%% file: 0_main.tex
\documentclass[%
 reprint,
groupedaddress,
 amsmath,amssymb,
 aps,
 prl,
]{revtex4-2}

\usepackage{graphicx}
\usepackage{dcolumn}
\usepackage{bm}

\usepackage{orcidlink}
\usepackage{multirow}
\usepackage{booktabs}
\input{defs}

\begin{document}

\preprint{APS/123-QED}

\title{Improved Limit on Neutrinoless Double Beta Decay of \mohundred~from AMoRE-I}

\input{author-orcid}

\date{\today}

\begin{abstract}
AMoRE searches for the neutrinoless double beta decay of  \mohundred{}
with 100 kg of enriched \mohundred{}.
Scintillating molybdate crystals coupled with a metallic magnetic calorimeter operate at milli-Kelvin temperatures to measure the energy of electrons emitted in the decay. 
AMoRE-I is a demonstrator for the full-scale AMoRE, operated at the Yangyang Underground Laboratory for over two years.
The exposure was 
8.02 kg$\cdot$year (or 3.89 kg$_{\mathrm{^{100}Mo}}\cdot$year),
and the total background rate near the $Q$ value was 0.025 $\pm$ 0.002 counts/keV/kg/year. We observed no indication of \znbb~decay and report a new lower limit of the half-life of~\mohundred~\znbb~decay as $ T^{0\nu}_{1/2}>2.9\times10^{24}~\mathrm{yr}$ at 90\% confidence level.
The effective Majorana mass limit range is \mbb$<$(210--610) meV
using nuclear matrix elements estimated in the framework of different models, including the recent shell model calculations.
\end{abstract}

\maketitle


Experiments with solar and atmospheric neutrinos \cite{Super-Kamiokande:1998kpq,SNO:2002tuh} have found that neutrinos are massive.
Various oscillation experiments \cite{RENO:2018dro,DayaBay:2018yms,NOvA:2021nfi,T2K:2023smv} have measured the three mixing angles and two mass differences.
Although the absolute masses of neutrinos have not yet been measured, we know they are very small, less than 1~eV/$c^{2}$ 
based on measurements of
the tritium $\beta$-spectrum endpoint~\cite{Katrin:2024tvg,Planck:2018vyg}. 

The small neutrino masses and the absence of right-handed neutrinos in the Standard Model motivated the introduction of Majorana masses for neutrinos, 
as opposed to charged leptons that have Dirac masses. 
Small Majorana neutrino mass 
can be generated via the seesaw mechanism 
in which the masses of active neutrinos are suppressed by heavy right-handed sterile neutrinos \cite{Mohapatra:1979ia}, and the mass terms violate lepton number conservation~\cite{Majorana:1937vz}. 

The currently well-established method for determining if neutrinos are Majorana fermions is to search for neutrinoless double beta $\rm (0\nu\beta\beta$) decay of nuclei \cite{Agostini:2022zub}.
Regardless of the underlying mechanism, the observation of $\rm 0\nu\beta\beta$ decay would prove that lepton number is violated \cite{Schechter:1981bd}, which is necessary for the seesaw mechanism and leptogenesis \cite{Fukugita:1986hr}.
The amplitude of \znbb~decay is proportional to the effective Majorana mass, defined by the charged-current couplings of Majorana neutrinos.
A \znbb~decay experiment requires monitoring a large mass of isotopes 
with a detector having an ultralow background and a high energy resolution to reach the true effective Majorana mass.
In spite of more than 
seventy years of experimental efforts \cite{Fireman:1949qs}, no signal for \znbb~decay has been observed.
The best half-life limit has been established by KamLAND-Zen for $\rm ^{136}Xe$ with $T_{1/2}^{0\nu} > 3.8 \times 10^{26}$~yr, limiting the effective Majorana mass to \mbb$<$(28--122)~meV \cite{KamLAND-Zen:2024eml}.
For other isotopes, the $T_{1/2}^{0\nu}$ (\mbb) limits are: $\rm 3.3 \times 10^{25}$~yr (90--305~meV) for $\rm ^{130}Te$ by CUORE using cryogenic techniques \cite{Campani:2024xsl}, $\rm 1.8 \times 10^{26}$~yr (79--180~meV) for $\rm ^{76}Ge$ by GERDA using high purity germanium 
detectors~\cite{GERDA:2020xhi}, and $\rm 1.8 \times 10^{24}$~yr (280--490 meV) for $\rm ^{100}Mo$ by CUPID-Mo using cryogenic scintillating crystal detectors \cite{CUPID:2020aow,Augier:2022znx}.

AMoRE aims to search for the $0\nu\beta\beta$~decays
using molybdate-based crystal detectors operating at milli-Kelvin (mK) temperatures \cite{Bhang:2012gn,AMoRE:2015asn}, similar to CUPID-Mo. 
AMoRE intends to achieve quasifree of background, defined as having less than one count in the region of interest (ROI) near $Q_{\beta\beta}$=3034.4 keV
for the five-year experiment with about 100 kg of $^{100}$Mo isotope. 

The AMoRE project is progressing in three phases. The first two, AMoRE-Pilot and AMoRE-I, were completed at the 700-meter-deep Yangyang Underground Laboratory (Y2L).
AMoRE-II will run at the newly built Yemilab for five years.
We have reported a half-life limit of $T_{1/2}^{0\nu}= 3.0\times10^{23}$~yr at a 90\% confidence level (CL) using six \enCMO~ crystals in the AMoRE-Pilot stage~\cite{Alenkov:2019jis,Alenkov:2021asw,AMoRE:2024cun}.
To confirm the performance and long-term stability of the detection system and determine the background level achievable with the existing setup at Y2L, we operated AMoRE-I from 2019 to 2023.
This report describes the experimental setup, data analysis, and a new half-life limit of \mohundred. 

AMoRE-I was carried out 
in the same cryostat 
that was used for AMoRE-Pilot~\cite{Alenkov:2019jis,Alenkov:2021asw,AMoRE:2024cun,Kang_2017},
but with a larger number of detectors. We made a few modifications to the detector modules and shielding enhancements. The AMoRE-I system used a two-stage temperature control system to maintain the detector tower at a constant temperature~\cite{Woo:2022qmh}. The datasets used for this report were taken at a temperature of 12\,mK. 

The AMoRE-I detector array consists of eighteen crystals with a total mass of 6.2~kg, including 3.0~kg of $^{100}$Mo, comprising five \enLMO~(LMO) crystals, and thirteen \enCMO~(CMO) crystals, six of which were inherited from AMoRE-Pilot.
The molybdenum in the crystals was enriched in $^{100}$Mo to 95.7$\pm$0.2\%.
Each detector module consisted of a crystal surrounded by a Vikuiti reflector film, an MMC sensor connected to the crystal, and a light detector made of an absorber with an MMC sensor. The mass-spring vibration damping system~\cite{Lee:2018pxi} was removed to accommodate the increased number of detector modules. 
The stainless steel screws in the detector modules were replaced with copper and brass 
for lower radioactive contamination. Additionally, Si wafers with SiO$_{2}$ antireflection coating were used for the light absorber instead of Ge wafers~\cite{mbkim2023ieeetas}. One flat surface of the LMO crystal, upon which the gold phonon collector was evaporated, was ground with 1500-grit SiC sandpaper to enhance the gold film bond at the interface. Each crystal had a stabilization heater attached to the flat surface to measure and correct the gain drift of the heat signal~\cite{dhkwon2020jltp}.
Pulses of fixed current and duration were injected through the heater every 10 seconds. Detailed descriptions of the detector module can be found in ~\cite{Kim:2022uce,Woo:2022qmh}.

An additional 5 cm of lead was added outside of the cryostat vacuum chamber to further reduce high-energy environmental $\gamma$'s. Initially, aluminum plates were used to support the lead bricks, but these plates were found to be highly contaminated 
by $\rm ^{228}Ra$ and were removed in May 2021. Ten additional muon counters made of plastic-scintillator panels were installed to extend the solid-angle coverage for 
the detector array.

Heat and light signals from the SQUID electronics were continuously digitized without an event trigger and stored using flash analog-to-digital converter modules with an 18-bit resolution for a 10-volt peak-to-peak dynamic range at a 100 kHz sampling frequency.
Data acquisition (DAQ) with the full eighteen-crystal detector array started in December 2020 and ended in April 2023, with a 93\% live time.
About 78\% of the DAQ live time was dedicated to physics measurement, while calibration and other commissioning data took up the remaining 
time.
Thorium-containing welding rods were placed between the outer vacuum chamber and the external lead shielding to calibrate the energy scale every 4--8 weeks, depending on the detector's stability.

\begin{figure*}[btph]
    \centering
    \includegraphics[width=\textwidth]{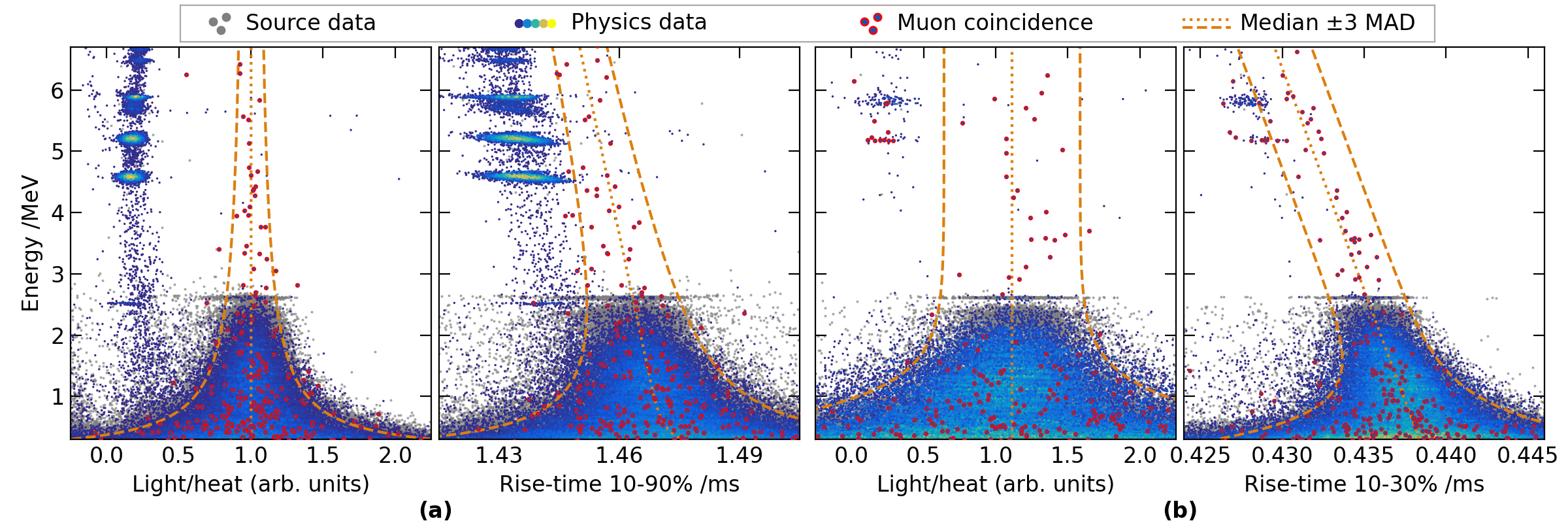}
    \caption{Particle discrimination parameters: light-to-heat ratios (L/H) and the raw heat signals' RTs of (a) a CMO and (b) an LMO detector. Dots with blue-yellow color gradients denote physics data, overlaid on the source data denoted as gray dots. Events in both 3-MAD bands for L/H and RT denoted as dashed-orange curves were selected as $\beta/\gamma$ events. Events in the muon veto window are indicated by red circles. Some $\alpha$-like events with muon coincidence at the electron equivalent energy slightly above 5 MeV in the LMO data are caused by the capture of muon-induced neutrons on the lithium-6 nuclei: $^{6}$Li($n,\alpha$)$t$.}
    \label{fig:PaID}
\end{figure*}

To suppress noise, the heat signal of each detector was processed
using a Butterworth bandpass filter for the offline analysis.
For most detectors, the trigger thresholds were set below 100\,keV.
Two modules that suffered from a significant vibration or electric noise had higher thresholds.
The amplitudes of the heat and light signals were determined by fitting template waveforms to the data using a least-squares fit~\cite{hslim2024}.
The fit was applied to the filtered waveforms
and the bandwidths of the Butterworth filters were optimized to provide the 
highest energy resolutions.
A template signal was constructed by averaging 2615 keV $\gamma$ events accumulated in calibration runs.
The rise and fall times were derived from the raw signals, defined as the time elapsed between specific points on the pulse.
Unlike the AMoRE-pilot analysis, which used the time difference between the 10\% and 90\% levels of the signal, this analysis defined rise time (RT) variably for each detector, optimizing the pulse shape discrimination (PSD) power.
One LMO detector had a noisy light signal, making it impossible to calculate the light signal amplitude accurately.
Data from this detector were excluded from the analysis because of the limited PSD power.
Examples of the RT versus signal amplitudes for a CMO and an LMO detector are shown in Fig~\ref{fig:PaID}.

The detector response change over time, mainly caused by temperature 
variations, influenced the pulse amplitudes and shapes of both physics and heater signals.
Correlations between the heater signal's RT and the amplitude of 2615 keV $\gamma$ events were determined with calibration data and used for drift correction in the corresponding dataset.
Each dataset consists of a calibration run followed by physics runs before the next calibration.

The $\beta/\gamma$ energy scale was initially calibrated using four prominent $\gamma$ peaks at 511.0 keV, 583.2 keV, 911.2 keV, and 2614.5 keV in the calibration-run spectrum for each dataset.
As in the previous analysis, signal amplitudes for each detector were all well described as a quadratic function of true energy values, without a constant term~\cite{KIM2017105,Alenkov:2019jis}.
Secondary energy calibration was performed using another quadratic function with a constant term, including more $\gamma$ peaks from the calibration data.

\begin{figure}[btp]
\centering
\includegraphics[width=0.48\textwidth]{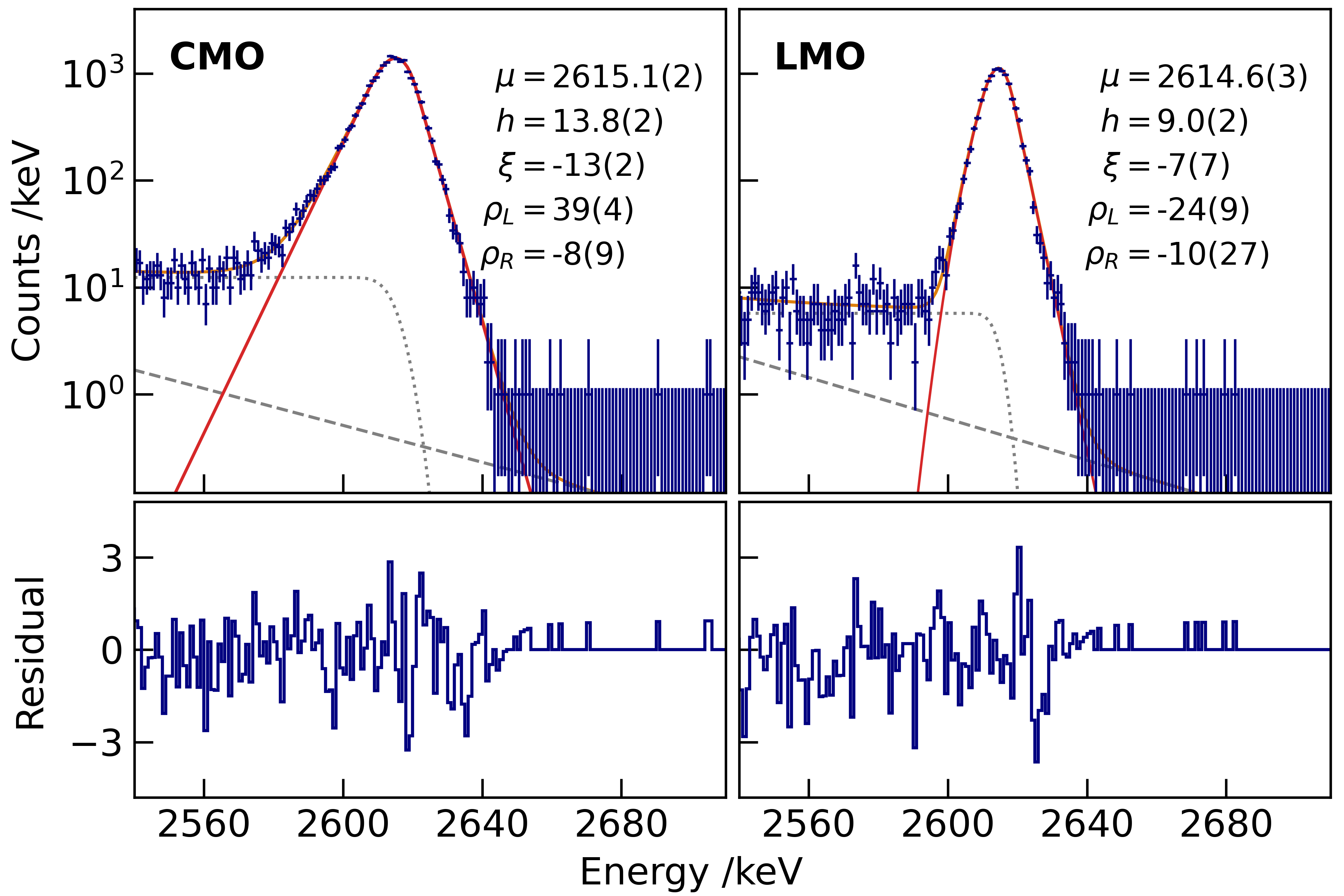}
\caption{
Part of the energy spectrum measured with the Th-containing sources (blue points with error bars) around 2615~keV $\gamma$ peaks for a CMO crystal (left) and an LMO crystal (right) are shown at the top. The fit function (solid-orange curve) consists of an exponential (dashed-gray curve) plus a smeared-step (dotted-gray curve) background component and a peak signal component that is represented by a Bukin function (solid-red curve). For the Bukin function, the peak location ($\mu$) and the full-width-at-half-maximum energy resolution ($h$) are given in keV units, and the asymmetry ($\xi$), left-tail ($\rho_{L}$), and right-tail ($\rho_{R}$) parameters are in percentages. The bottom panels show the fit residuals defined for the data bins with positive counts as $([\mathrm{data}]-[\mathrm{fit}])/\sqrt{[\mathrm{data}]}$.}
\label{fig:peak_shape}
\end{figure}

Owing to nonuniform, position-dependent detector responses, peaks in the energy spectra were asymmetric, showing tails on both the lower and upper sides.
Among various peak-fitting functions, the Bukin function~\cite{bukinfunction} was found to be well suited for the peaks in the calibration spectrum. The spectra around the 2615 keV peak in the calibration data for a CMO and an LMO detector are shown in Fig.~\ref{fig:peak_shape}.
Shape parameters that were determined from fitting below 2615 keV were extrapolated to $Q_{\beta\beta}$ to estimate the 
\znbb{} signal shape.
The squares of energy resolutions ($h^{2}$) were well fitted to a quadratic function of energy.
The energy resolution at $Q_{\beta\beta}$ differs for each detector, varying from 10 to 28 keV of full width at half maximum with a weighted average 13 keV.
Other shape parameters for asymmetry ($\xi$), left tail ($\rho_{L}$), and right tail ($\rho_{R}$) varied slowly with energy, 
and their uncertainties were extrapolated using a linear function or were left constant, depending on each detector's characteristics.

The \znbb{} event-selection criteria for background reduction were 
divided into two categories.
The first category was particle identification to remove the continuous $\alpha$ background around the ROI due to radioactive contamination on the crystal surface or surrounding materials.
Two parameters were adopted for each detector: PSD using RT of the heat signal, and the light-to-heat ratio (L/H), which leverages the differences in scintillation quenching for $\alpha$ and $\beta/\gamma$ particles.
Fig.~\ref{fig:PaID} shows the RT and L/H versus energy for the calibration runs, background runs, and events that were coincident with muons for one CMO and one LMO detector. Here, the $\alpha$ events are clustered at higher energies and at smaller RT and L/H values, while most of the $\beta/\gamma$ events without muon coincidence lie below 2.6 MeV.

The median-absolute-deviation (MAD) values were determined using source data at the energy 
range of 465--2665 keV and fitted using a function of energy. Events within the 3 MAD bands for RT and L/H, denoted as the dashed-orange curves in Fig.~\ref{fig:PaID} were selected as $\beta/\gamma$ events.
The selection efficiencies at the calibration $\gamma$-peak energies were determined by comparing the event counts at each peak before and after the selection using the Bukin function on top of a simple exponential or linear background to extract the peak from the resulting spectra.
These $\beta/\gamma$ selection efficiencies, when extrapolated to $Q_{\beta\beta}$, varied among detectors and ranged between 80\% and 95\% with uncertainties on the 1\% level.
Generally, CMO detectors showed a much better discrimination power than LMO, for both RT and L/H parameters, due to the higher light yield and the correlation of light yield with the pulse shape~\cite{Mailyan:2023qpa,krwoo2024jltp}.
The alpha backgrounds for the six CMO detectors used in AMoRE-Pilot were reported in \cite{Alenkov:2021asw}. 
The activities of $\rm ^{226}Ra$ ($\rm ^{228}Th$) for the AMoRE-I dataset were determined by analysis of $\alpha-\alpha$ coincident events from the sequential decay of $^{222}$Rn and $^{218}$Po ($^{224}$Ra and $^{220}$Rn). These were 5--59 $\mu$Bq/kg (1--13 $\mu$Bq/kg) in the CMO detectors, except for one highly contaminated CMO crystal,
and 1--2 $\mu$Bq/kg for both $\rm ^{226}Ra$ and $\rm ^{228}Th$ in the LMO detectors.
Detailed background modeling of AMoRE-I is in progress. 

The second event-selection category involved a series of anticoincidence cuts.
First, events occurring within 10\,ms after a muon candidate event were rejected. About ten thousand muon candidate events were registered daily, resulting in an inefficiency of only 0.1\% with negligible uncertainty.
Secondly, we imposed a single-hit cut in which the signal candidate is the only hit that occurred within a 2\,ms time window.
Since the trigger rate of the background runs was typically $\sim$1 Hz, the efficiency of this single-hit requirement was about 99.8\%.
The final anticoincidence condition, called $\alpha$ tagging,
targeted one of the major backgrounds in the ROI from the $\beta$ decay of $^{208}$Tl with $Q\sim$4.998~MeV. These background events follow an $\alpha$ event emitted in $^{212}$Bi decay to $^{208}$Tl.
Since the half-life of $^{208}$Tl is 3.053 minutes and the energy released in the $\alpha$ decay of $^{212}$Bi is 6.207\,MeV, all events within 20 minutes after an $\alpha$ event  in the same crystal with 6.2 $\pm$ 0.1\,MeV energy were rejected.
Two detectors with exceptionally high $\alpha$ rates had a narrower $\alpha$ energy window of 40~keV~\cite{Alenkov:2021asw}.
The efficiency of this $\alpha$ veto was about 98\% on average: 78\% for the most $\alpha$-contaminated CMO and higher than 99\% for LMOs. The cut efficiencies are summarized in Table \ref{tab:efficiency}. 

\begin{figure*}[btp]
\includegraphics[width=0.99\textwidth]{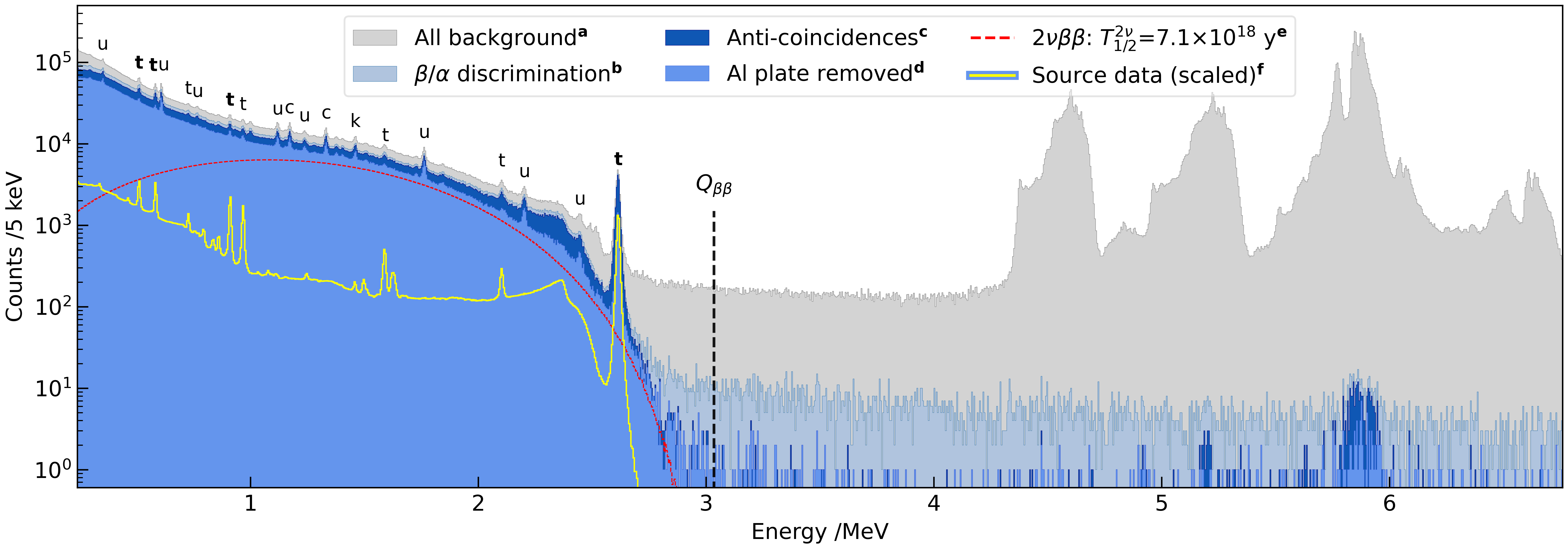}
\caption{Energy distributions of AMoRE-I data corresponding to 8.02 kg$\cdot$year exposure, following step-by-step selections.
a: raw physics events, 
b: $\beta/\gamma$ events after selection using the pulse shape discrimination and L/H ratio,
c: after all anticoincidence selections, 
d: after removal of the aluminum plates,
e: simulated two-neutrino double beta ($2\nu\beta\beta$) decay component corresponding to the exposure of {\bf d}~\cite{GEANT4:2002zbu,Ponkratenko:2000um,CUPID-Mo:2023lru}.
The source calibration spectrum ({\bf f}, yellow line) is scaled to match the size of 2615~keV $\gamma$ peak in {\bf d}.
Some background events remained in the high-energy range after all cuts, such as those in 5--6 MeVee, due to inefficient $\alpha$ rejection in the LMO data.
Letters above prominent $\gamma$ peaks denote the corresponding decay chains: u for uranium-238, t for thorium-232, c for cobalt-60, and k for potassium-40.
}
\label{fig:result_spectrum}
\end{figure*}
\begin{table}[tbp]
    \caption{
    Efficiencies of the selection cuts utilized to suppress background in the vicinity of the expected \znbb{} peak, and the typical systematic uncertainties. The efficiencies averaged over all the detectors are given in parentheses. The \znbb~ containment efficiency is calculated using the DECAY-0 event generator~\cite{Ponkratenko:2000um} and the GEANT4-based detector simulation~\cite{GEANT4:2002zbu}.
    }\label{tab:efficiency}
    \begin{tabular}{@{ }llcc@{ }}
    \toprule
    \multicolumn{2}{l}{Parameters/selection} & Efficiency (\%) & Uncertainty (\%) \\
    \midrule
    \multicolumn{2}{l}{Particle identification} & &\\
    \hspace{.5cm} & PSD$\times$L/H (3 MAD)& 78.8--95.4 (89.9) & 1.6 \\
    \multicolumn{2}{l}{Anti-coincidences} & &\\
    & Multiplicity ($\mathcal{M}=1$) & 99.8 & $<$0.1 \\
    & Muon veto (10 ms) & 99.9 & $<$0.1\\
    & $^{212}$Bi $\alpha$ veto (20 min) & 77.6--99.8 (97.8) & 0.6 \\
    \multicolumn{2}{l}{\znbb~containment} & 78.4--82.4 (81.1) & 1.0 \\\midrule
    \multicolumn{2}{l}{Total detection efficiency}  & 49.1--76.1 (70.9) & 1.6 \\\bottomrule
    \end{tabular}
\end{table}

Figure~\ref{fig:result_spectrum} shows the resulting background energy spectra accumulated for 8.02 kg$\cdot$year (or 3.89 kg$_{\mathrm{^{100}Mo}}\cdot$year) exposure, following the step-by-step selections, overlaid with the calibration spectrum.
The first dataset taken with the aluminum support plate, shown as the dark blue shaded area, represents about 14\% of the total exposure but contains about 20\% of the events in the energy range below 2.7 MeV, including about 60\% of 2615~keV $\gamma$ events. Despite this, the background due to the contaminated aluminum plate is mainly confined to energies below 2.7 MeV, so these data were included in the \znbb{} analysis.
As shown in the previous studies~\cite{Alenkov:2019jis}, the relationship between the signal amplitude and energy differs for $\alpha$ events compared to $\beta/\gamma$ events due to pulse shape differences, which vary across detectors. 
Consequently, $\alpha$ energies, which were determined separately, are not reported in this work.
As a result, the energy spectrum for all detectors and datasets shown in Fig.~\ref{fig:result_spectrum} has many $\alpha$ peaks above 4 MeV electron-equivalent energy that are distributed incoherently when they are calibrated with the functions for $\beta/\gamma$ events.

Events in the 2.7--3.6 MeV energy range were selected for the \znbb{} study.
The background energy distribution was approximated as a linear combination of flat and exponential background components.
The exponential component includes high-energy $2\nu\beta\beta$-decay events and $\beta$ background from internal and surface contamination of the crystal~\cite{Alenkov:2021asw}. The flat component is introduced to describe high energy $\gamma$s from neutron capture, external sources such as rock and radon in the air,
and residual $\beta$ and $\alpha$ events due to incomplete rejection.
A more detailed background modeling based on the measurements of detector components' radioactivities is in progress.

Considering the background ($b$) and the \znbb{} decay signal ($s$) can be expressed in terms of the decay rate ($\Gamma^{0\nu}$=$\ln 2/T^{0\nu}_{1/2}$), the number of $^{100}$Mo nuclei 
($N_{100}$),
the detection efficiency ($\varepsilon$) for the \znbb~decay of \mohundred~(shown in Table \ref{tab:efficiency}), and the DAQ live time ($t$) as $s$=$\varepsilon\Gamma^{0\nu}N_{100}t$.
An unbinned likelihood function was constructed as follows:
\begin{equation}
\mathcal{L}=\prod_{i=1}^{N}\prod_{j=1}^{n_{i}}\frac{(s_{i}+b_{i})^{n_{i}}\,e^{-s_{i}-b_{i}}}{n_{i}!}\cdot f^{s+b}_{i}(E^{i}_j)\cdot\pi(\varepsilon_{i}, \bm{\nu}_{i}),
\label{eq:likelihood}
\end{equation}
where $i$ and $j$ are indices of the detector and its events, respectively, and $n_{i}$ is the number of observed events in the $i$th detector.
The expected spectral shape was described by the probability density function, $f^{s+b}$, that includes a background model, with flat and exponential components, plus a signal peak in the form of a Bukin function for fully contained $0\nu\beta\beta$ events. 
Efficiency ($\varepsilon_{i}$) and peak shape parameters ($\bm{\nu}_{i}$) were treated as nuisance parameters with Gaussian priors ($\pi$).

\begin{figure}[ht]
    \centering
     \includegraphics[width=0.49\textwidth]{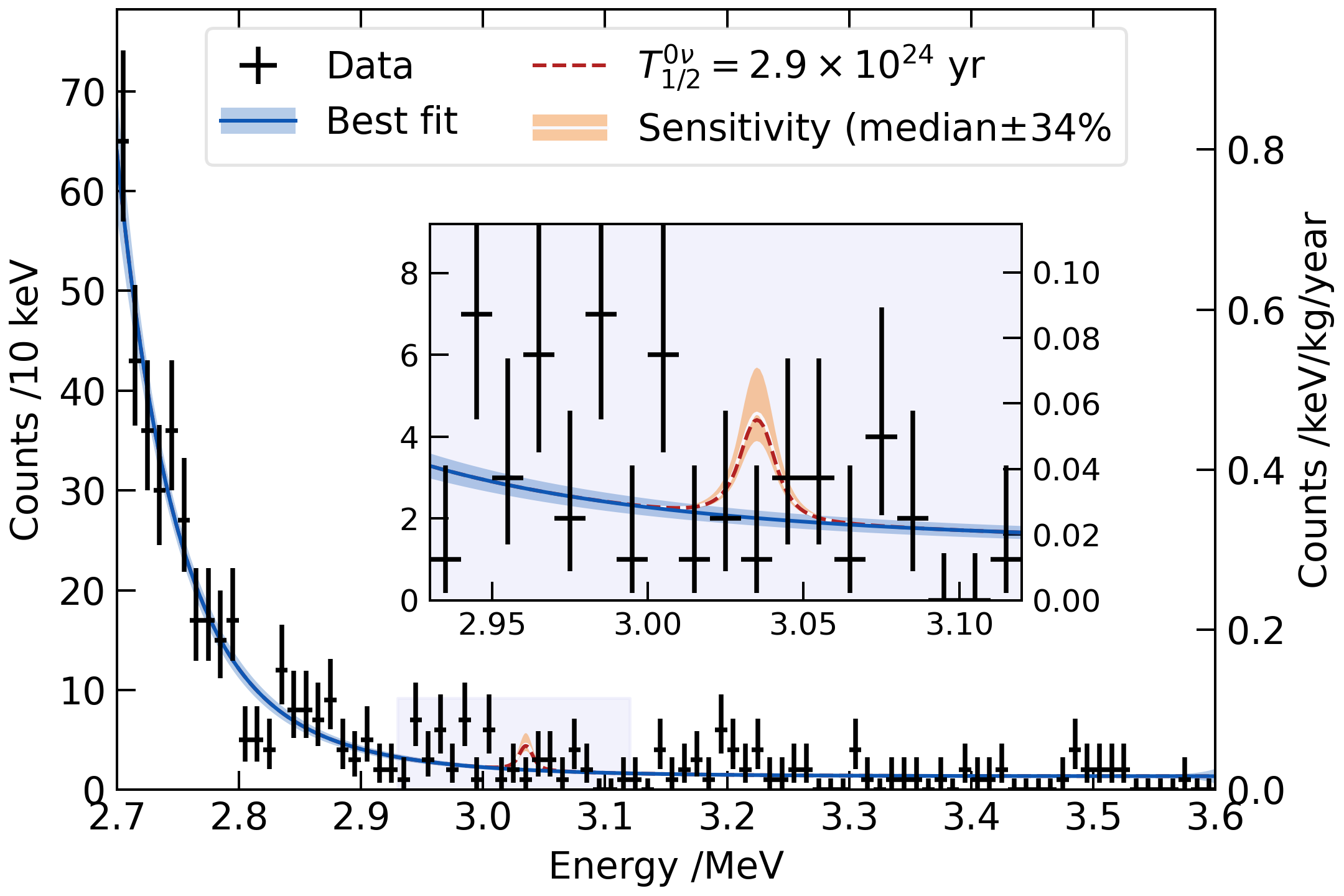}
    \caption{The energy spectrum of selected events in the region of interest. The points with error bars are measured data with the $1\sigma$ Poisson confidence intervals. 
    The solid blue curve is the best fit with a null signal, with the shaded region denoting the combined uncertainty of the fit. The dashed red curve shows the expected signal shape for a 0$\nu\beta\beta$ decay half-life of 2.9$\times10^{24}$ yr, the upper limit at a 90\% CL from this study. The white curve with the orange shaded area shows the median and $\pm$34\% sensitivity for the 90\% CL limit.} 
    \label{fig:sig90peak}
\end{figure}
The minimized negative log likelihood was profiled over the sensitive range of $\Gamma^{0\nu}$ from 0 to $\sim10^{-23}$ year$^{-1}$, and the best fit was found at $\Gamma^{0\nu}=0.0$ year$^{-1}$, meaning that no event excess was found over the assumed background shape, as shown in Fig. \ref{fig:sig90peak}. 
The corresponding limit on the half-life of the \znbb{} decay of $^{100}$Mo was evaluated at a 90\% CL to be:
\begin{equation}
T^{0\nu}_{1/2}>2.9\times10^{24}~\mathrm{yr},
\label{halfliferesult}
\end{equation} 
which extends the CUPID-Mo limit~\cite{Augier:2022znx}.
The limit matched well to the sensitivity determined by numbers of pseudodatasets with the given exposure, peak shapes and sideband background levels.
The median $\pm$34\% sensitivity of 90\% CL limit is 2.8$^{+1.0}_{-0.9} \times 10^{24}$ yr, as shown in Fig.~\ref{fig:sig90peak}.

The posterior analysis
leads to a total background level 
of $b=0.025\pm0.002$ counts/keV/kg/yr.
The LMO detectors showed a slightly lower background rate of $0.021\pm0.005$ counts/keV/kg/yr on average, while 
the background counting rate of the CMO detectors is $0.026\pm0.003$ counts/keV/kg/yr.
The total background rate in AMoRE-I was reduced by $\sim$15 times compared to that of AMoRE-pilot.

\begin{figure}
    \centering
    \includegraphics[width=0.48\textwidth]{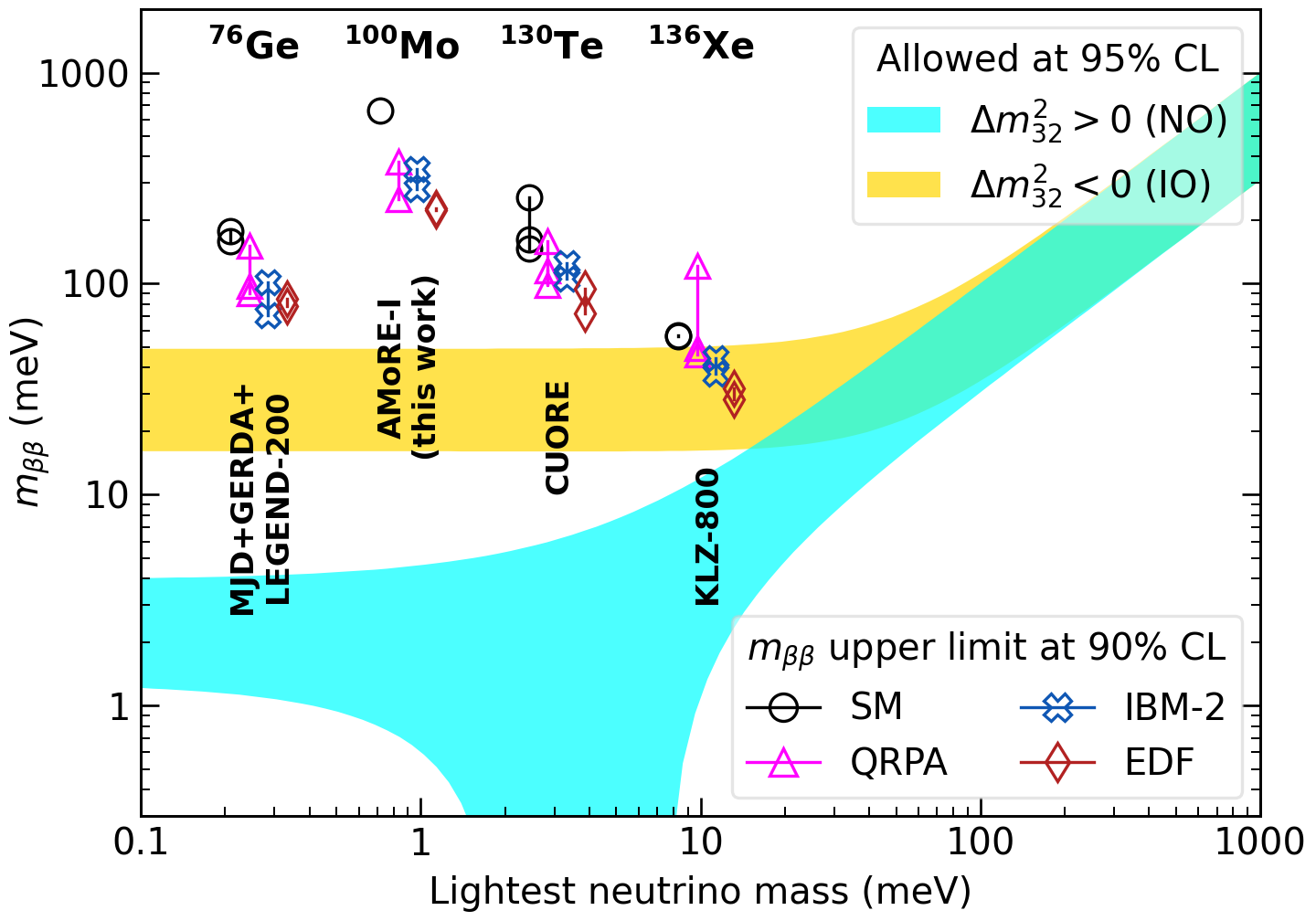}
    \caption{Data points denote upper limits on the effective Majorana neutrino mass ($m_{\beta\beta}$) by experiments~\cite{LEGEND:2024nu,CUORE:2024ikf,KamLAND-Zen:2024eml} and by AMoRE-I, using nucluear matrix element caculations~\cite{Rath:2013fma,Simkovic:2013qiy,LopezVaquero:2013yji,Barea:2015kwa,Hyvarinen:2015bda,Song:2017ktj,Simkovic:2018hiq,Rath:2019fsp,Coraggio:2022vgy}: SM=shell model, IBM-2=interacting boson model-2, QRPA=quasiparticle random phase approximation, EDF=energy-density functional theory. Underlaid colored areas area allowed region in the two-dimensional space of $m_{\beta\beta}$ and the lightest neutrino mass by the neutrino oscillation experiments at 95\% confidence level.}
    \label{fig:mbb}
\end{figure}

We calculated the effective Majorana mass within the theoretical framework of the light neutrino exchange model, incorporating the phase space factor \cite{Kotila:2012zza,Mirea:2015nsl,Stoica:2019ajg} and nuclear matrix elements (NMEs) \cite{Rath:2013fma,Simkovic:2013qiy,LopezVaquero:2013yji,Barea:2015kwa,Hyvarinen:2015bda,Song:2017ktj,Simkovic:2018hiq,Rath:2019fsp}.
The range of inferred upper limits on the effective Majorana mass is \mbb$<$(210--350) meV assuming an axial vector coupling constant $g_A=1.27$. 
If we include the first shell model calculation recently published for \mohundred~\cite{Coraggio:2022vgy},
the range of limits extends to \mbb$<$ (210--610) meV.
The lower limit is derived from an energy density functional considering the nuclear deformation and pairing fluctuation~\cite{LopezVaquero:2013yji}, while the upper limit is based on a shell model that explicitly includes the short-range correlations~\cite{Coraggio:2022vgy}.
Upper limits on the effective Majorana mass from existing experimental data and NMEs are shown in Fig.~\ref{fig:mbb}.

AMoRE-II is under preparation at Yemilab \cite{Park:2024sio}
with a muon rate about a quarter of that at Y2L~\cite{Prihtiadi:2017sxc}.
We have developed an LMO detector module with improved energy resolution and alpha background rejection \cite{Kim:2022qgl,Kim:2022ubc}.
The background level for AMoRE-II is projected to be less than $10^{-4}$ counts/keV/kg/yr based on radioassay data and GEANT4 simulations~\cite{backgroundamore}.
The discovery sensitivity is projected to be approximately $\rm 4.5\times 10^{26}$ yr for five years of data collection. 

\emph{Acknowledgements}--This research is supported by Grants No. IBS-R016-D1 and No. IBS-R016-A2.
It is also supported by the National Research Foundation of Korea (NRF-2021R1I1A3041453, NRF-2021R1I1A6A1A03043957) and the National Research Facilities \& Equipment Center (NFEC) of Korea (No. 2019R1A6C1010027).
We appreciate the support by the Ministry of Science and Higher Education of the Russian Federation (N121031700314-5), the MEPhI Program Priority 2030.
The group from the Institute for Nuclear Research (Kyiv, Ukraine) was supported in part by the National Research Foundation of Ukraine under Grant No. 2023.03/0213.
We thank the Korea Hydro and Nuclear Power (KHNP) Company for providing underground laboratory space at Yangyang and the IBS Research Solution Center (RSC) for providing high-performance computing resources.
These acknowledgements are not to be interpreted as an endorsement of any statement made by any of our institutes, funding agencies, governments, or their representatives.
\appendix

\bibliography{reference}

\end{document}

%% file: defs.tex
\newcommand{\enCMO}{$^{\mathrm{48depl}}$Ca$^{100}$MoO$_{4}$}

\newcommand{\enLMO}{$\rm Li_{2}{}^{100}MoO_{4}$}
\newcommand{\mbb}{$ m_{\beta\beta}$}

\newcommand{\znbb}{$0\nu\beta\beta$}

\newcommand{\mohundred}{$\rm ^{100}Mo$}

%% file: author-orcid.tex
\author{A.~Agrawal\,\orcidlink{0000-0001-7740-5637}}
\author{V.V.~Alenkov\,\orcidlink{0009-0008-8839-0010}}
\author{P.~Aryal\,\orcidlink{0000-0003-4955-6838}}
\author{J.~Beyer\,\orcidlink{0000-0001-9343-0728}}
\author{B.~Bhandari\,\orcidlink{0009-0009-7710-6202}}
\author{R.S.~Boiko\,\orcidlink{0000-0001-7017-8793}}
\author{K.~Boonin\,\orcidlink{0000-0003-4757-7926}}
\author{O.~Buzanov\,\orcidlink{0000-0002-7532-5710}}
\author{C.R.~Byeon\,\orcidlink{0009-0002-6567-5925}}
\author{N.~Chanthima\,\orcidlink{0009-0003-7774-8367}}
% ...
\collaboration{AMoRE~Collaboration}
\noaffiliation